\documentclass[preprint,aps,nofootinbib,preprintnumbers,amsmath,amssymb]{revtex4} 
\usepackage{graphicx}
\usepackage{dcolumn}
\usepackage{bm}
\usepackage{amssymb}
\usepackage{latexsym}
\usepackage[colorlinks, linkcolor=blue, citecolor=blue, urlcolor=blue]{hyperref}
\newcommand{\be}{\begin{equation}}
\newcommand{\ee}{\end{equation}}
\newcommand{\bq}{\begin{eqnarray}}
\newcommand{\eq}{\end{eqnarray}}

\begin{document}

\title{Dark radiation from a unified dark fluid model}

\author{Chao-Qiang Geng$^{1,2,3}$\footnote{geng@phys.nthu.edu.tw},
Lu-Hsing Tsai$^{2}$\footnote{lhtsai@phys.nthu.edu.tw}, and Xin Zhang$^{4,5}$\footnote{zhangxin@mail.neu.edu.cn}}
\affiliation{\\$^{1}$College of Mathematics \& Physics, Chongqing University of Posts \& Telecommunications, Chongqing, 400065, China
\\$^{2}$Department of Physics, National Tsing Hua University, Hsinchu, Taiwan
\\$^{3}$Physics Division, National Center for Theoretical Sciences, Hsinchu, Taiwan
\\$^{4}$Department of Physics, College of Sciences, Northeastern University, Shenyang 110004, China
\\$^{5}$Center for High Energy Physics, Peking University, Beijing 100080, China}

\begin{abstract}
We present  a unified dark fluid model to describe the possible evolutionary behavior 
of $\Delta N_\mathrm{eff}$ in dark radiation. This model can be viewed as an interacting 
model for the dark sectors, in which dark matter interacts with dark radiation. We show 
that the evolution of $\Delta N_\mathrm{eff}$ can be nicely explained without some
drawbacks, such as the blowup of $\Delta N_\mathrm{eff}$ and the 
non-vanishing interaction at the late time.
\end{abstract}

\maketitle

\section{Introduction}

The $\Lambda$CDM model has successfully explained many
important cosmological observations such as the acceleration of the universe
and the radial velocity distribution of the galaxies as well as the cosmic
microwave background (CMB) fluctuations~\cite{WMAP9,Planck2013}. Besides
the motivation of the theoretical completeness, from the viewpoint
of the observational data there still leaves some room for the
existence of physics beyond $\Lambda$CDM. Recently, the analysis of
the pure CMB data shows that the effective number of relativistic
degrees of freedom is $N_\mathrm{eff}=3.36^{+0.68}_{-0.64}$ (95\%
CL)~\cite{Planck2013}, which accommodates the standard model (SM)
prediction of $N_\mathrm{eff}^\mathrm{SM}=3.046$~\cite{NeffSM} within
$1\sigma$ range, while the combined analysis with the measurement of
$H_0$ gives $N_\mathrm{eff}=3.62^{+0.50}_{-0.48}$ (95\%
CL)~\cite{Planck2013}, which is larger than the SM value
at around $2\sigma$ level. The extra degree of freedom is usually referred to as dark
radiation (DR). It is worth noticing that the extra radiating
component can be extracted by the probe of the primordial deuterium and helium abundances 
at the big bang nucleosynthesis (BBN) epoch~\cite{BBN0}. For instance, 
it has been recently shown that 
$N_\mathrm{eff}=3.71^{+0.47}_{-0.45}$ and $3.50\pm0.20$
in Refs.~\cite{BBN1} and \cite{BBN2},
respectively.

Many models have been used to describe $\Delta N_\mathrm{eff}\equiv
N_\mathrm{eff}-N_\mathrm{eff}^\mathrm{SM}$. Among them, 
imposing a new relativistic degree of freedom
beyond the SM is a straightforward way~\cite{PPPV},
but such a scenario can only explain the case in which DR is in an
equal amount at the BBN and CMB epoches, namely, $N_{\rm eff}^{\rm
BBN}=N_{\rm eff}^{\rm CMB}$. 
Note that
there may be a tension  between BBN and CMB for $N_{\rm eff}$ as the current data seems to indicate
that $N_{\rm eff}^{\rm
CMB}<N_{\rm eff}^{\rm BBN}$.
In order to understand such a decrease (or increase) of
$N_{\rm eff}$ at CMB, various subtle models have been proposed in which some
interactions between DR and dark matter (DM) are assumed. For example,
if heavy DM particles\footnote{For the ultra light DM candidates, 
some interesting properties were discussed in Ref.~\cite{LightDM}.}
 can decay into relativistic states, the
increase in DR could be interpreted; see, e.g.,
Refs.~\cite{ParticleDecay0,pd0,ParticleDecay,modeldep} for model dependent and
independent analyses.

Since there is no evidence that the dark sectors are independent to
each other, an interaction between DM and DR is quite possible. 
Models related to this possibility have been widely discussed in the 
literature~\cite{IM:1,IM:AC,IM:2,IM:3}. However, it should be pointed out that there
are still some drawbacks in these models: some of them blow up $\Delta N_\mathrm{eff}$ 
in the late time, which is also equivalent to the existence of non-vanishing interaction between 
DM and DR in the present.
In this paper, we propose a unified dark fluid model describing both
DM and DR, which can nicely yield the decrease (or increase) in $\Delta
N_\mathrm{eff}$ without the above drawbacks.

This paper is organized as follows. In Sec.~\ref{sec:2}, we introduce the
unified dark fluid model. In
Sec.~\ref{sec:3}, we discuss the extra effective relativistic degree of freedom
$\Delta N_\mathrm{eff}$. Conclusions are given in
Sec.~\ref{sec:4}.

\section{A unified dark fluid}\label{sec:2}

We start with a dark fluid, in which the energy density is expressed as
\begin{eqnarray}
\rho_{\rm
dark}=(Aa^{-4(1+\alpha)}+Ba^{-3(1+\alpha)})^{1/(1+\alpha)}\;,\label{CtEq}
\end{eqnarray}
where $a$ is the scale factor of the universe, $\alpha$ is a small
real number, and $A$ and $B$ are positive, which can be determined by the
initial condition at some specific time. Note that this dark fluid can be
viewed as a mixture of DM and DR. In fact, it is a special case
of the new generalized Chaplygin gas (NGCG) model with the equation of 
state (EOS) $w=1/3$ proposed in Ref.~\cite{Zhang:2004gc}. We remark that this 
model is also inspired by the generalized Chaplygin gas (GCG)
scenario which unifies DM and dark energy (in the case of
the cosmological constant) in a single fluid~\cite{GCG}. For
$\alpha=0$ in Eq.~(\ref{CtEq}), the energy density reduces to the
sum of matter and radiation forms. For $\alpha$ being a small real
number, the fluid can exhibit the behavior of both matter and
radiation.

  From the continuity equation,
$\dot\rho+3H(\rho+P)=0$, the pressure of the dark fluid $P_{\rm
dark}$ can be derived, and then the EOS
parameter of the dark fluid can be obtained to be
\begin{eqnarray}
w_\mathrm{dark}={P_{\rm dark}\over \rho_{\rm
dark}}={Aa^{-4(1+\alpha)}\over
3(Aa^{-4(1+\alpha)}+Ba^{-3(1+\alpha)})}\;.
\end{eqnarray}
For a small $a$, we have $w_\mathrm{dark}\simeq 1/3$, which is the same as 
the radiation fluid, while for a large $a$, the fluid behaves like
matter with $w_\mathrm{dark}\simeq 0$. Similar to GCG, this
fluid can be naturally decomposed into two interacting components
with constant EOS parameters, $w=1/3$ and $w=0$, respectively. As a result,
this unified dark fluid model can also be regarded as an interacting
dark-sector model in which DM interacts with DR.

Subsequently, we can write $\rho_{\rm dark}=\rho_{\rm
dm}+\rho_{\rm dr}$ and $P_{\rm dark}=P_{\rm dm}+P_{\rm dr}$.
By using $P_{\rm dm}=0$ and $P_{\rm
dr}=(1/3)\rho_{\rm dr}$, we derive
\begin{eqnarray}\label{EqDensityCom}
\rho_\mathrm{dm}&=&K^{1\over
1+\alpha}\left(1-{Aa^{-4(1+\alpha)}\over
K}\right)\,,\,~~~\rho_\mathrm{dr}=K^{1\over
1+\alpha}{Aa^{-4(1+\alpha)}\over K}\;,
\end{eqnarray}
where $K\equiv Aa^{-4(1+\alpha)}+Ba^{-3(1+\alpha)}$. Evidently, $A$
and $B$ can be naturally determined by the initial condition of the
two components, e.g., the DR and DM densities at the present time.

The energy transfer from DM to DR in unit
volume and in unit time can be derived as
\begin{eqnarray}
Q&=&-3\alpha H{P_{\rm dark}\over\rho_{\rm dark}}(\rho_{\rm
dark}-3P_{\rm dark})=-\alpha
H{\rho_\mathrm{dr}\rho_\mathrm{dm}\over\rho_\mathrm{dm}+\rho_\mathrm{dr}}\;,
\end{eqnarray}
where the sign of
$\alpha$ fixes the direction of the energy flow. A positive $\alpha$
makes the energy flow from DR to DM, whereas the negative one reverses the
direction. By this definition, the energy
continuity equations for DM and DR are given by
$\dot{\rho}_{\rm dm}+3H\rho_{\rm dm}=-Q$ and $\dot{\rho}_{\rm
dr}+4H\rho_{\rm dr}=+Q$, respectively. Note that if
$\rho_\mathrm{dr}\gg(\ll)\rho_\mathrm{dm}$, the energy transfer $Q$ can
be reduced to $Q=-\alpha H \rho_{\mathrm{dm}(\mathrm{dr})}$. 
This kind of the interaction simultaneously
involves the two important forms $Q=-\alpha H \rho_\mathrm{dm}$ and
$-\alpha H \rho_\mathrm{dr}$, studied extensively in the
interacting dark energy models~\cite{GCGratio}. These
forms, similar to those obtained from the GCG fluid, are crucial features of 
the GCG-like model~\cite{GCG}. We remark that once $Q$ is proportional to the Hubble expansion
rate $H$, there is a factor of $T^2$ in the radiation dominated
epoch. We will discuss the effect of the interactions on the
time evolution of $\Delta N_\mathrm{eff}$ in the next section.

It should be pointed out that
although our model is inspired by the GCG and NGCG models, 
there are some significant differences between the DM-DR interacting model and the 
DM-dark energy interacting model, in particular when the cosmological perturbations are considered.
For example, for the GCG model, when it is considered as a 
unified model the perturbation calculations force it to be extremely close to the $\Lambda$CDM model 
($\alpha<10^{-6}$)~\cite{tegmark02,park09}, whereas
 a much wider range of $\alpha$ is allowed, 
i.e., $\alpha$ may be of the order ${\cal O}(10^{-1})$~\cite{wangyt13},
when it is treated as a model of vacuum energy interacting with DM. 
The case of the NGCG model is discussed in Ref.~\cite{zx13}. 
The primary cause is that dark energy is a non-adiabatic fluid so that how to treat its pressure perturbation is 
obscure to some extent.\footnote{In the case of dark energy, since $w$ is negative, its adiabatic sound speed $c_a$ 
would be imaginary due to $c_a^2=dp_{de}/d\rho_{de}=w$ (for the example of constant $w$), leading to instability in the theory. In order to fix this problem, 
it is necessary to assume that dark energy is a non-adiabatic fluid and impose a physical sound speed $c_s^2>0$ by hand. 
Usually, $c_s$ is set to be the light speed as if the dark energy fluid is realized by a scalar field, which is what is done in the 
{\tt CAMB} and {\tt CMBFAST} codes. But such a treatment would also lead to some instabilities, 
in particular for the $w=-1$-crossing models and some specific interacting dark energy models.
For more detailed discussions, see Ref.~\cite{zxPPF14}.} 
Nevertheless, for the model considered in this paper, since both DM and DR are adiabatic fluids, 
 our model can be treated as a model of unified dark fluid as well as a model of DM interacting with DR. 
As a result, we
expect that the constraints on our model from ``geometry measurements'' and ``structure's growth measurements'' 
will be consistent owing to the fact that both DM and DR are adiabatic fluids with
well defined sound speeds   and well treated pressure perturbations.

\section{$N_\mathrm{eff}$ in interacting models}\label{sec:3}

From the definition of $N_\mathrm{eff}$, the extra relativistic
energy density exceeding the $\Lambda$CDM model is given by
\begin{eqnarray}
\Delta \rho_\nu\equiv\Delta N_\mathrm{eff}{7\over8}\bigg({T_\nu\over
T_\gamma}\bigg)^{4}\rho_\gamma^0a^{-4}\;,\label{rhonu}
\end{eqnarray}
where $({T_\nu/T_\gamma})=(4/11)^{1/3}$ after the photon was
heated at the $e^+e^-$ annihilation epoch.

On terms of the description of our model, DR interacts with DM, 
which results in the deviation $\Delta \rho_\nu$ from the standard evolution behavior in the
$\Lambda$CDM model. Consequently, if we identify $\rho_{\rm dr}$ with
$\Delta \rho_\nu$ in Eq.~(\ref{rhonu}), we obtain a time
evolutionary $\Delta N_\mathrm{eff}$. On the other hand, since DM
also differs from the standard scaling $a^{-3}$, we can write
$\rho_{\rm dm}=\rho_{\rm dm}^0f(a)a^{-3}$, where $f(a)$ represents the departure 
from the standard result. The
explicit form of $f(a)$ can be extracted from
Eq.~(\ref{EqDensityCom}). Hence, according to the decomposition of
the model, we have
\begin{eqnarray}
\rho_\mathrm{dark}=\rho_\mathrm{dr}+\rho_\mathrm{dm}
&=&\rho_{\rm dark}^0\Big(ra^{-4(1+\alpha)}+(1-r)a^{-3(1+\alpha)}\Big)^{1/(1+\alpha)}\nonumber\\
&=&\Delta N_\mathrm{eff}(a){7\over8}\bigg({T_\nu\over
T_\gamma}\bigg)^{4}\rho_\gamma^0 a^{-4}+\rho_\mathrm{dm}^0 f(a)
a^{-3},
\end{eqnarray}
where $\rho_{\rm dark}$ has been re-parameterized
by the value $\rho_{\rm dark}^0$ at the present time and a
dimensionless parameter $r$, taken around
$10^{-5}$, which is of the same order as the radiation fractional
density now.

\begin{figure}
\centering
\includegraphics[width=8cm]{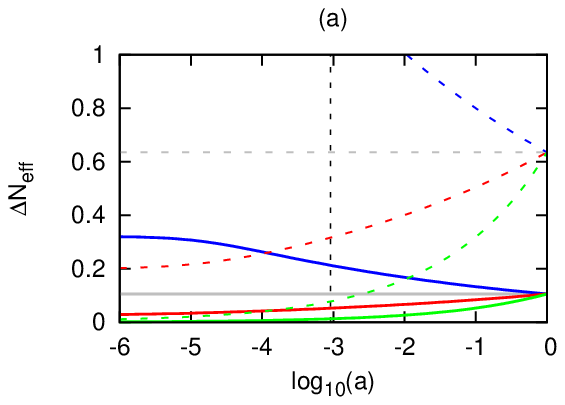}
\includegraphics[width=8cm]{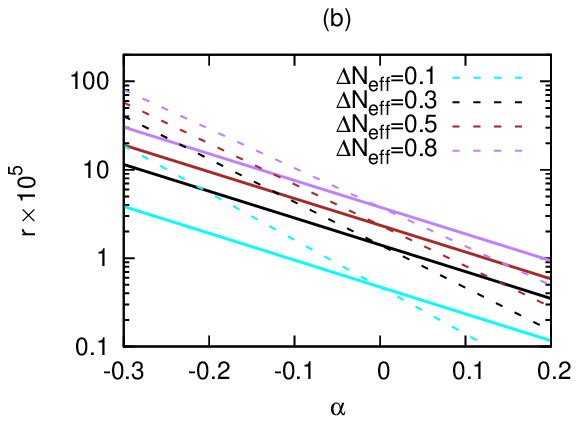}
\caption{(a) $\Delta N_\mathrm{eff}$ versus the scale factor $a$, where
the blue, red, green, gray curves represent $
\alpha=0.1$, $-0.1$, $-0.3$, and $0$, and solid and
dashed ones denote $r=0.5\times 10^{-5}$ and $3\times 10^{-5}$, respectively, while
the black dashed line indicates the scale corresponding to the CMB epoch.
(b) Correlations between $r$ and $\alpha$, where the cyan, black, brown, and
purple curves stand for $\Delta N_\mathrm{eff}=0.1$, $0.3$, $0.5$, and $0.8$,
while solid and dashed ones correspond to the CMB and BBN epochs,
respectively.}\label{Fig_Neff}
\end{figure}

In Fig.~\ref{Fig_Neff}a, we show $\Delta N_\mathrm{eff}$ as a
function of the scale factor $a$ for $\alpha=0.1$ (blue), $-0.1$ (red),
$-0.3$ (green), and $0$ (black), and $r=0.5\times 10^{-5}$ (solid) and $3\times 10^{-5}$ (dashed),
respectively. All curves with $r$ fixed approach the
same $\Delta N_\mathrm{eff}$ at $a=1$, which are sensitive to the initial condition 
$\rho_{\rm dark}^0$ for $a\gtrsim 10^{-2}$. For $\alpha=0.1$, $\Delta N_\mathrm{eff}$ is a decreasing function, 
while for $\alpha=-0.1$ or $\alpha=-0.3$, it
behaves as an increasing one. Notice that $\alpha=0$ gives the constant value of $\Delta N_\mathrm{eff}$
 due to the vanishing of interacting term $Q$. Different choices of $r$ will
lead to different results. From the figure, it is clear that $\alpha>0$ with the
energy flowing from DM to DR is favored. In Fig.~\ref{Fig_Neff}b,
we illustrate the correlations between the two parameters $r$ and $\alpha$
with different choices of $\Delta N_\mathrm{eff}$ in the CMB and BBN
epochs, respectively. 
In the figure, the contours with
the cyan, black, brown, and
purple curves stand for $\Delta N_\mathrm{eff}=0.1$, $0.3$, $0.5$, and $0.8$,
while solid and dashed ones correspond to $z=1100$ and $10^{-9}$, respectively.
From Fig.~\ref{Fig_Neff}b, we can roughly determine the
model parameters. For example, if we assume $\Delta
N_{\rm eff}^{\rm CMB}=0.3$ and $\Delta N_{\rm eff}^{\rm BBN}=0.5$,
which are consistent with the current observations, we obtain $\alpha\simeq0.15$
and $r\simeq 0.5\times 10^{-5}$, which are reasonable parameters for
the model. Note that a positive value of $\alpha$ is
required if the result of the smaller $\Delta N_\mathrm{eff}$ in CMB
persists in the future observations.

\begin{figure}
\centering
\includegraphics[width=8cm]{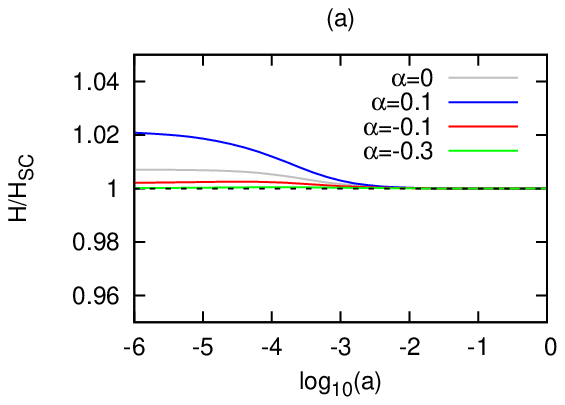}
\includegraphics[width=8cm]{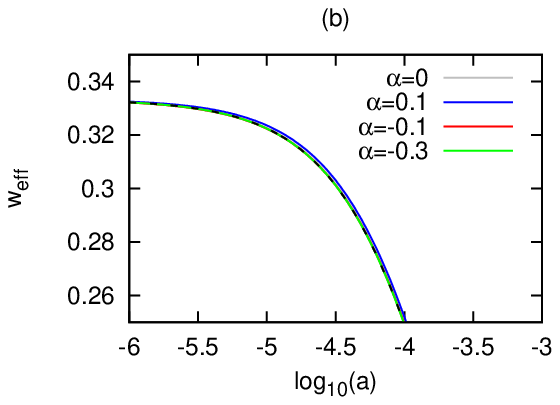}
\caption{(a) $H/H_\mathrm{SC}$ and (b) $w_\mathrm{eff}$ versus the scale factor $a$
with $r=0.5\times 10^{-5}$, where the blue, red, green, and gray curves represent $\alpha=0.1$, $-0.1$,
$-0.3$, and $0$, respectively, while the black dashed curve in (b) represents  
  $w_{\rm eff}$ in the $\Lambda$CDM model.}\label{Fig_Hubble}
\end{figure}

\begin{figure}
\centering
\includegraphics[width=10cm]{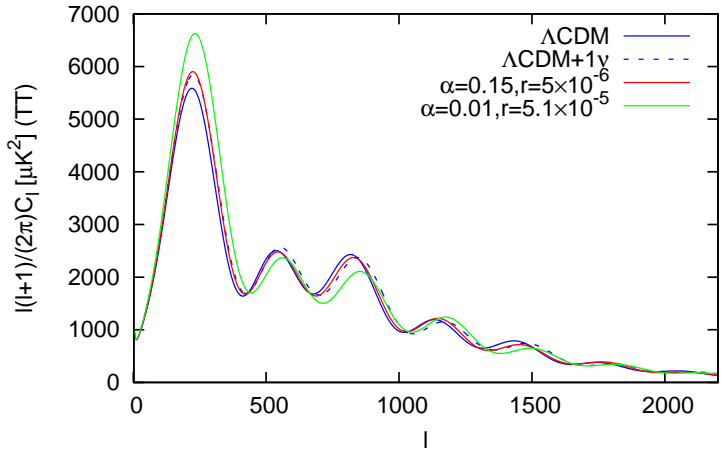}
\caption{The CMB temperature power spectrum in models, where the upper to lower (blue solid, dashed, red and green)
 curves correspond to 
$\Lambda$CDM, $\Lambda$CDM with one massless sterile neutrino, $(\alpha,r)=(0.15,5\times 10^{-6})$ and $(0.01, 5.1\times 10^{-5})$,  respectively.}\label{Fig_PowerSpec}
\end{figure}

In Fig.~\ref{Fig_Hubble}a, we plot the ratio of $H/H_\mathrm{SC}$ as a function of the scale factor $a$,
 where $H$ and $H_\mathrm{SC}$ are the Hubble parameter in our model and $\Lambda$CDM, respectively. 
 The parameters are taken as
 $r=0.5\times 10^{-5}$, and  $\alpha=0.1$ (blue), $-0.1$ (red), $0$ (black), and
$-0.3$ (green) for the plots.
  It is shown that the cosmic expansion becomes faster in the early time
   for any value of  $\alpha$. It can be easily understood since  $\Delta N_{\rm eff}$ is always positive
   as shown in Fig.~\ref{Fig_Neff}a, which implies the existence of extra energy density apart from that given by 
   $\Lambda$CDM. In addition, the EOS parameter $w_\mathrm{eff}$ versus $a$
is given in Fig.~\ref{Fig_Hubble}b. In the figure, we also show the result (dashed curve) for $\Lambda$CDM. 
It is worth noting that the evolution of the Hubble parameter $H$ could also provide some effect 
on the anisotropic CMB power spectrum. The increasing of $H$ at the CMB epoch ($\log_{10} a \simeq -3$) 
for any values of $\alpha$ would not only suppress the damping tail to equivalently solve the anomaly of DR, 
but also shift the acoustic peak slightly toward a smaller angular scale (larger $\ell$), while the value of 
the first acoustic peak could be lifted up. Typically, by taking $r\simeq 5\times10^{-6}$ and $\alpha=0.15$, 
 the first peak could rise about the same amount as that in the scenario of adding an additional massless sterile neutrino 
 into $\Lambda {\rm CDM}$. For a larger $r$, the amplitude of the power spectrum increases rapidly. 
 These effects are illustrated in Fig.~\ref{Fig_PowerSpec}, where
 several models including $\Lambda {\rm CDM}$, $\Lambda {\rm CDM}$ with an extra massless sterile neutrino, and 
 our unified fluid model with $(\alpha,r)=(0.15,5\times 10^{-6})$ and $(0.01, 5.1\times 10^{-5})$ are presented.
 A similar discussion on the anisotropic spectrum for other interacting models was also given in Ref.~\cite{IM:1}.

To illustrate our results, we now compare our model with other two interacting models, 
{\em Models A} and {\em B}, in which  the energy transfers are $Q_A=\alpha_1 H
\rho_\mathrm{dm}$ and $Q_B=\rho_\mathrm{dm}/\tau_{\rm dm}$, respectively, with $\alpha_1$ and $\tau_{\rm dm}$ 
being the free parameters. {\em Model A} is a simple
interacting scenario between DM and DR, which is studied in Ref.~\cite{IM:1}, 
while {\em Model B} is examined in Refs.~\cite{ParticleDecay0,pd0},
in which  the interacting term $\rho_\mathrm{dm}/\tau_{\rm dm}$ 
can be directly interpreted as the energy transferring  
into the DR component from the decaying of heavy particles 
with the life time $\tau_{\rm dm}$,  around the BBN epoch. 
Unlike our model and {\em Model A}, the energy density $\rho_{\rm dm}$ for a heavy particle in {\em Model B} 
is unlikely to be linked with DM due to the short life time $\tau_{\rm dm}$ of only a few orders of seconds~\cite{pd0}.

\begin{figure}
\centering
\includegraphics[width=8cm]{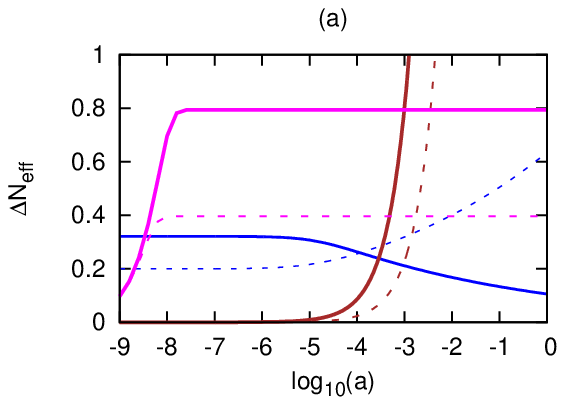}
\includegraphics[width=8cm]{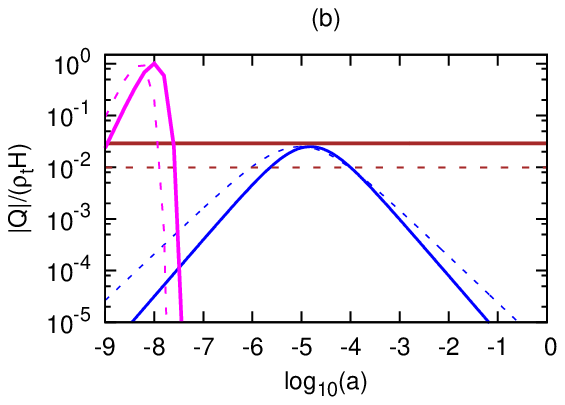}
\caption{ Evolutions of (a) $\Delta N_\mathrm{eff}$ and (b) $q\equiv |Q|/(\rho_{\rm t} H)$,
where  blue solid (dashed), brown solid (dashed), and  magenta solid (dashed) curves represent
$\alpha=0.1\;(-0.1)$ and $r=0.5\times10^{-5}\;(3\times 10^{-5})$ in our model, $\alpha_1=0.03\;(0.01)$ in {\em Model A},
and   $\tau_{\rm dm}=2000\,{\rm s}\;(500\,{\rm s})$ and $\rho_{\rm dm}/s=2\times 10^{-3}{\rm MeV}$ in {\em Model B}, respectively.
}\label{Fig_Com}
\end{figure}

In Fig.~\ref{Fig_Com}a, we present $\Delta N_\mathrm{eff}$ as a function of the scale factor $a$ 
in different models. For {\em Model A (B)}, we will use $\alpha_1=0.03$ and $0.01$ 
($\tau_{\rm dm}=2000\,{\rm s} \mbox{ and } 500\,{\rm s} $) as input parameters.  
$\rho_{\rm dm}^0$ in {\em Model A} is identical to the DM density in the present, 
while for {\em Model B} we will fix the  comoving energy density 
$(\rho_{\rm dm}/s)=(2\times 10^{-3}){\rm MeV}$ at BBN, with $s$ being the entropy density at that time. 
In both models $\rho_{\rm dr}\times a^4$ at very early time is taken to be zero as the other initial condition. 
We see that in {\em Model A}, $\Delta N_\mathrm{eff}$ coincides with the
observation at the CMB era, but it blows up in the late time.
In {\em Model B}, $\Delta N_\mathrm{eff}$ only increases at very early time and behaves 
as a constant after $a\gtrsim 10^{-8}$. The average rate of the change in $\Delta N_{\rm eff}$ 
from BBN to CMB for our model is faster than {\em Model B}, but gently than {\em Model A}. 
Moreover, both increasing and decreasing behaviors of $N_{\rm eff}$ can be described in our model, 
which could be a potential target for probing this model in the future observations. 
In addition, the dimensionless relative energy transfer $q\equiv |Q|/(\rho_{\rm t} H)$ with
$\rho_{\rm t}$ being the sum of energy densities of DM and DR is
plotted in Fig.~\ref{Fig_Com}b for each case. With the same
parameter values in Fig.~\ref{Fig_Com}a, $q$ in Model A always behaves as a constant 
due to the crucial feature $(\rho_{\rm dr}/\rho_{\rm dm})\simeq\alpha$~\cite{IM:1}, 
whereas in the late time the nonzero value of $q$ indicates that the
interaction between DM and DR is still rather strong even at present. 
In {\em Model B}, the region of the nonzero $q$ centralizes at the beginning of BBN with the order of magnitude 
around the peak as large as order unity. In our model, $|Q|/H$ is proportional to $\rho_\mathrm{dm}$ and
$\rho_\mathrm{dr}$ in very early and late times, respectively,
so that a nonzero value of $q$ can only be confined in some range of time.
Obviously, the behaviors of $\Delta N_\mathrm{eff}$ and $q$ in our
model are more reasonable than {\em  Model A}.

\section{Conclusions}\label{sec:4}

We have proposed a unified dark fluid model to understand 
the possible evolutionary behavior of $\Delta N_\mathrm{eff}$ in DR.
Inspired by the GCG model, the dark fluid can be viewed as a scheme for 
the unification of DM and DR. Such a fluid behaves like radiation and matter 
in the radiation and matter dominated epochs, respectively.
Interestingly, this model can also be regarded as an interacting model in the dark sectors
as DM interacts with DR with the form explicitly obtained.
Moreover, we have evaluated  the evolution of $\Delta
N_\mathrm{eff}$ in DR, which is favored by the current
observational data for $\alpha>0$. Comparisons with the other two interacting
models, $Q=\alpha_1 \rho_\mathrm{dm} H$ and
$\rho_\mathrm{dm}/\tau_{\rm dm}$, have been also given. 
We have shown
that our predicted values of $\Delta N_\mathrm{eff}$ and $q$ in the unified
dark fluid model are more reasonable than {\em Model A}. 
In particular, in our model there are no drawbacks, such as the blowup of $\Delta N_\mathrm{eff}$ 
and the non-vanishing interaction at the late time.
Clearly, more accurate analyses on $N_{\rm eff}$ and its evolution  in the future
could help to identify if our model is a viable scenario.

\begin{acknowledgments}
This work was supported by the National Center for Theoretical
Sciences, National Science Council (Grant Nos.
NSC-98-2112-M-007-008-MY3 and NSC-101-2112-M-007-006-MY3) and
National Tsing-Hua University (Grant Nos. 102N1087E1 and 102N2725E1) at Taiwan,
R.O.C, as well as by the National Natural Science Foundation of
China (Grant Nos. 10705041, 10975032 and 11175042) and the National
Ministry of Education of China (Grant Nos.~NCET-09-0276, N100505001
and N120505003).
\end{acknowledgments}

\end{document}